\documentclass[12pt,a4paper]{article}
\pdfoutput=1
\usepackage{jheppub}
\usepackage{amsthm}

\usepackage{mathtools}
\DeclarePairedDelimiter\abs{\lvert}{\rvert}

\usepackage{color,soul}

\usepackage{graphicx}
\usepackage{feynmp}

\newtheorem*{con}{Conjecture}
\newtheorem{thm}{Theorem}
\newtheorem{cor}{Corollary}

\def\sgn{\text{sgn}}

\def\[{\left[}
\def\]{\right]}
\def\({\left(}
\def\){\right)}

\newcommand{\be}{\beta}

\def\sgn {\text{sgn}}

\newcommand*\pFq[2]{{}_{#1}F_{#2}\genfrac[]{0pt}{}}

\def\tr{\mathrm{tr}}
\def\Tr{\mathrm{Tr}}

\def\sgn{\text{sgn}}

\def\tr{\mathrm{tr}}
\def\Tr{\mathrm{Tr}}

\def \be {\begin{equation}}
\def \ee {\end{equation}}
\def \bea {\begin{eqnarray}}
\def \eea {\end{eqnarray}}
\def \beal#1 {\begin{align}#1\end{align}}

\everymath{\displaystyle}

\preprint{TIFR/TH/19-26}

\title{Global Symmetry and Maximal Chaos }

 \author[a,1]{Indranil Halder\note{indranil.halder@tifr.res.in}  }

\affiliation[a]{Department of Theoretical Physics, Tata Institute of Fundamental Research, Homi Bhabha Rd, Mumbai 400005, India}

\abstract{In this note we study chaos in generic quantum systems with a global  symmetry generalizing seminal work [arXiv : 1503.01409] by Maldacena, Shenker and Stanford. We conjecture a bound on instantaneous chaos exponent in a thermodynamic ensemble at temperature $T$ and chemical potential $\mu$ for the continuous global symmetry under consideration. 
For local operators which could create excitation up to some fixed charge, the bound on chaos exponent is independent of chemical potential $\lambda_L \leq \frac{2 \pi T}{ \hbar} $.
On the other hand when the operators could create excitation of arbitrary high charge, we find that exponent must satisfy  $\lambda_L \leq \frac{2 \pi T}{(1-\abs{\frac{\mu}{\mu_c}}) \hbar} $, where $\mu_c$ is the maximum value of chemical potential for which the thermodynamic ensemble makes sense.
 As specific examples of quantum mechanical systems we consider conformal field theories. In a generic conformal field theory with internal $U(1)$  symmetry living on a  cylinder the former bound is applicable, whereas in more interesting examples of holographic two dimensional conformal field theories dual to Einstein gravity, we argue that later bound is saturated in presence of a non-zero chemical potential for rotation. }

\begin{document}

\maketitle

\section{Introduction}

There is a deep relation between chaos and thermal physics. It is generically believed that chaotic behaviour of trajectories in phase-space is a way to mixing and thermalisation.\footnote{A system might or might not be ergodic on whole energy surface. Classically for a weakly perturbed integrable system with with given number of degrees of freedom (for example,
harmonic oscillators with small nonlinear couplings) according to the Kolmogorov-Arnold-Moser theorem phase space is foliated by invariant tori almost everywhere,
and we do not expect it to be ergodic on whole energy subspace. On the other hand a system of particles interacting through a hard-sphere potential with perfectly reflecting boundary conditions in a box is known to be fully chaotic, with no invariant tori in the phase space. For systems with additional (extensive) conserved quantities there exists a more general notion of thermalisation restricted to a subspace of energy surface determined by those conserved quatities.  } Chaos is aperiodic long-term behaviour that exhibits sensitive dependence to initial conditions. Classically a measure of sensitivity is 
 \begin{equation}\label{sensitive}
 	\begin{aligned}
 		\frac{\partial q(t)}{\partial q(0)}=\{q(t),p(0)\} 
 	\end{aligned}
 \end{equation}
 On the other hand, thinking about chaos in terms of states in quantum mechanics is bit tricky since  time evolution in quantum mechanics is unitary and linear, and so two states that are close together in the sense of having a large inner product will remain so. In this context semi-classical notion of chaos is to be understood by the fact that states that are orthogonal can still be physically similar - in the course of time this is not preserved \cite{Peres:1984zz, Feingold:1984d}.
Instead of prospective of states we will take an operator viewpoint and focus on generalizing \eqref{sensitive} by replacing Poisson brackets by commutators and averaging over thermal ensemble. Actually a better quantity to consider is squared commutator (to avoid cancellation due to phases) of two approximately local hermitian operators  (which increases energy of the ensemble by an amount of thermal energy scale $\beta^{-1}$ (temperature) and has vanishing one point function)  \cite{Larkin}
\begin{equation}\label{CC}
 	\begin{aligned}
 		C(t)=-\langle [W(t),V(0)]^2  \rangle
 	\end{aligned}
 \end{equation}
Before we proceed to discuss behaviour of the quantity defined in \eqref{CC} in a chaotic system as a function of time we take a moment to define an important timescale relevant for chaos, the exponential decay time $t_d$ for two point functions like $\langle X(t) X(0) \rangle$. This is also the time scale after which time-ordered correlation functions attend their large time asymptotic. In strongly coupled interacting systems $t_d $ is same order as thermal scale $\beta$. Quantum mechanically butterfly effect can be defined as the phenomena that $C(t)$ becomes order $2 \langle V(0)V(0) \rangle \langle W(0)W(0) \rangle$ for times comparable to scrambling time $t_*$ for generic operators $V,W$.  We will restrict our attention to theories such that $t_*>>t_d$ (we present examples of systems satisfying this criterion in next paragraph). The typical behaviour of $C(t)$ as a function of time in chaotic systems is as follows. For times $t_*>t>t_d$, $C(t)$ grows exponentially like $ \epsilon e^{\lambda_L t}$ ($\epsilon$ is a system dependent constant), exponent $\lambda_L$ is called Lyapunov exponent. For times larger than $t_*$ eventually $C(t)$ saturates to above mentioned value as it thermalizes and the physics around this is governed by exponents dictated by Ruelle resonances.\footnote{In the context of holographic conformal field theories, Ruelle exponents gets related to the physics of quasi-normal modes in the bulk:  $ \langle W(t)W(0) \rangle \sim e^{-i \omega t}$  for $t$ times larger than decay times $t_d$ which is determined by (the smallest) imaginary part of quasi-normal mode frequency. As a result time ordered correction functions like $\langle V(0)V(0)W(t)W(t) \rangle$ is well approximate by $\langle V(0)V(0) \rangle \langle W(0)W(0) \rangle$ for $t>t_d$. On the other hand Lyapunov growth is related to the fact that  a boundary time defines a frame in the bulk, and relative to this frame, the quanta released at a very early time will have exponentially blue-shifted local energy determined by the boundary time difference. As a result when we are looking at out of time ordered quantities like $\langle W(t)V(0)W(t)V(0) \rangle$  quanta created by $W,V$ particles undergo a high-energy almost elastic collision for times in between decay time and scrambling time $t_*>t>t_d$ (here we are considering the situation with a  non-zero impact parameter). For times larger than scrambling time this correction function is dominated by small momentum modes and again decay is determined by  quasi-normal mode of $V,W$ with smallest imaginary part.  } 

Now we review examples displaying chaotic dynamics in the canonical ensemble. In semi-classical limit $\hbar \to 0$ intuition from classical dynamics suggests $t_d $ is order as $ \lambda_L^{-1}$ (collision time) and $t_* = \lambda_L^{-1} \log{\hbar^{-2}}$ (Ehrenfest time). For the rest of this paper we will not be working in semi-classical regime and we set $\hbar=1$ henceforth. Another important class of examples where Lyapunov exponent has been computed consists  of large $N$ field theories. In this context large hierarchy between scrambling and dissipation scales is controlled by $\log N^2$. For small values of 't Hooft coupling $\lambda$ the theory remains weakly coupled.  In the context of matrix models (with $\phi^4$ interaction) chaos exponent is determined in \cite{Stanford:2015owe}.\footnote{This is done by summing over a suitable infinite class of Feynman diagrams generalizing BFKL technique \cite{Kuraev:1977fs}.} Conclusively it is determined by interactions at energies of the order of thermal mass $m_{th}$ of the excitation (instead of energies at thermal scale $\beta^{-1}$ which is taken to be much larger, i.e., $\lambda^2<<m_{th} \beta <<1$ for the computation) dictating a very small chaos exponent proportional to $\lambda^2$ as expected for weakly coupled field theories \cite{Sekino:2008he}. In the weak coupling regime vector models (with $\phi^4$ interaction) also feature small chaos exponent in a rather different way, chaos exponent is order $1/N$ suppressed compared to matrix model result stated above \cite{Chowdhury:2017jzb}.\footnote{Weak coupling calculations of chaos exponent are performed in certain gauged models as well. In $QED_3$ with $N_F$  fermionic matter fields chaos exponent is $1/N_F$ suppressed in leading order in large $N_F$ expansion \cite{Steinberg:2019uqb}. Another context is gauged maximally supersymmetric matrix quantum mechanics in $0+1$ dimensions in the large $N$ limit. Here chaos exponent is proportional to dimensionless 't Hooft coupling as computed in \cite{Gur-Ari:2015rcq}. } On the other hand for large values of 't Hooft coupling the theory is strongly coupled and these are  studied using bulk/boundary weak/strong duality. Those conformal field theories which can be holographically described by Einstein gravity, the methods\footnote{Based on calculation of time delay of a probe particle in shock wave geometry created by the other operator in the bulk or equivalently by studying Regge limit of CFT correlation function.}  of \cite{Shenker:2013pqa, Shenker:2013yza, Roberts:2014isa, Roberts:2014ifa, Jackson:2014nla} give 
\begin{equation}\label{MaxChaos}
 \begin{aligned}
  \lambda_L=\frac{2\pi}{\beta }
 \end{aligned}
\end{equation}
Corrections to this result, when the bulk theory dual to thermal field theory is described by weakly coupled string theory with large radius of curvature, is considered in \cite{Shenker:2014cwa} and that results in decreases of chaos exponent.\footnote{It is given by $\lambda_L=\frac{2\pi}{\beta } \left( 1-\frac{1}{2}M^2 l_s^2+.. \right)$. Here $l_s$ is string scale and $M^2$ is a  positive number and we are considering spherical or planar horizons and $M$ determines transverse profile $h$ of the shock wave in this geometry $ (\nabla^2_{\perp}-M^2)h=0$ } In $T\bar{T}$ deformed quantum mechanics chaos exponent has been studied in \cite{Gross:2019ach} reporting the exponent given in \eqref{MaxChaos}.

In \cite{Maldacena:2015waa} it has been conjectured that in canonical ensemble for any unitary casual quantum mechanical system maximal chaos exponent is given by \eqref{MaxChaos}.\footnote{Curiously in disorder averaged (not unitary in strict sense)  quantum mechanical systems like SYK model chaos exponent is computed in \cite{Kitaev, Maldacena:2016hyu} featuring the same qualitative phenomena of small chaos exponent in weak coupling and maximal chaos exponent given by \eqref{MaxChaos} in strong coupling.  On the other hand their unitary cousin so called tensor models  may not feature chaos in strong coupling (to the leading order in large $N$) due to presence of large number of  additional non-chaotic soft modes (see \cite{Choudhury:2017tax}) compared to SYK like models mentioned above. }   The Argument reaching to this conclusion is based on analytic nature   of the  out-of-time-order correlation  function $F(t)=\langle W(t)V(0)W(t)V(0) \rangle$ (it is present when we expand the commutator in \eqref{CC}) in the following domain
\begin{equation}\label{OTOC}
	\begin{aligned}
		D= ( t: t_*>\operatorname{Re}{(t)}>t_d,\ \  \beta/4>\operatorname{Im}{(t)}>-\beta/4)
	\end{aligned}
\end{equation}
Rest of the proof relies on use of maximum modulus principle and Schwarz-Pick lemma on the function $F$ to show there exits a bound on $\lambda_L$ and the bound  is inversely proportional to width of the strip.\footnote{Existence of similar bound based on eigenstate thermalization hypothesis (ETH) is presented in \cite{Murthy:2019fgs}.} In this note we will be concerned with quantum mechanical systems with a global symmetry and in thermodynamic ensemble where chemical potential $\mu$ for the global symmetry turned on. 
It is important to note that although the  Hamiltonian is a positive definite operator, the charge operator for the conserved symmetry is not. For all the states contributing to the partition function we get exponential decay from positivity of energy whereas charge can contribute to exponential growth as well.
Therefore if the Hilbert space is such that in large energy sector we have an finite number of states whose charge is of the same order as energy, partition function is analytic only only for $\mu<\mu_c$ for some critical chemical potential $\mu_c$. The same principle dictates domain of analyticity of \eqref{OTOC} sinks (for operators which can create excitation of arbitrarily large charge) and 
as a result we find a chemical potential dependent  bound on the time derivative of the out of time ordered correlation function $F$. Under certain simplifying assumptions instantaneous chaos exponent for such a general ensemble satisfies\footnote{To be precise we emphasize that time averaged value of out of time ordered correction function is not considered here and that might in general satisfy a stronger constraint.} 
\begin{equation}\label{TheBound}
 \begin{aligned}
  \lambda_L \leq \frac{2\pi}{ \beta \left(1- \left| \frac{\mu}{\mu_c} \right| \right)} 
 \end{aligned}
\end{equation}
The formula \eqref{TheBound} is expected to be valid for small values of chemical potential for times near scrambling time. This is the central result of this note.  Naively as we approach the critical value of the chemical potential the bound becomes weaker and weaker, on the other hand the time period for which we could saturate the bound presented in \eqref{TheBound} is determined by smallness of chemical potential and therefore we don't expect the formula in \eqref{TheBound} to be valid for chemical potential near its critical value. Nevertheless understanding physics of the phenomena near critical chemical potential would be extremely interesting. 

The paper is organised as follows. In subsection \ref{statement} we state the conjecture precisely followed by an argument leading to it in subsection \ref{argument}. Then in section \ref{examples} we present examples of quantum mechanical systems where the bound \eqref{TheBound} is saturated. Finally in section \ref{remarks} we present some thoughts on future directions of development to conclude.

\section{The conjecture}

\subsection{Statement}\label{statement}

We consider a quantum mechanical system with a (continuous, $0$-form) global symmetry. The global symmetry generator is the hermitian operator $Q$ which commutes with the  Hamiltonian $H$ of the system $ [H,Q]=0 $. We turn on (real) chemical potential $\mu$ for this global symmetry and study thermodynamics of the system at inverse temperature $\beta=\frac{1}{T}$. The partition function of the system is 
\begin{equation}\label{partitionFun1}
	\begin{aligned}
		Z= \Tr ( e^{-\beta H +\beta \mu Q})
	\end{aligned}
\end{equation}
A complete set of common eigenstates of $H,Q$ is given by\footnote{Here we are hiding degeneracy labels  for notational simplicity.}
\begin{equation}
	\begin{aligned}
		& H | \phi_n \rangle=E_n | \phi_n \rangle, ~~~ Q | \phi_n \rangle =Q_n | \phi_n \rangle \\
		& \sum_n | \phi_n \rangle \langle \phi_n |=1, ~~ \langle \phi_m,\phi_n \rangle =\delta_{m,n}
	\end{aligned}
\end{equation}
In the basis of these eigen states
\begin{equation}\label{partitionFun2}
	\begin{aligned}
		Z= \sum_n e^{-\beta E_n +\beta \mu Q_n}
	\end{aligned}
\end{equation}
Suppose at asymptotically large values of energy, the spectrum is such that
\begin{equation}\label{condition}
	\begin{aligned}
		|Q_n|=\frac{E_n}{\mu_c}+\text{ sub-leading terms in $E_n \to \infty$ limit}
	\end{aligned}
\end{equation}
then summation in \eqref{partitionFun2} diverges for $\mu > \mu_c$.\footnote{Here we are making the reasonable assumption that density of states in large energies is not exponential in energy.} So the thermodynamic ensemble makes sense for chemical potential up to $\mu_c$.\footnote{The fact that presence of chemical potential may change domain of analyticity is a familiar one. For example see \cite{Raju:2018xue}.  } 

The central quantity of interest for us is the out of time ordered correlation function of two hermitian operators $V,W$ of the form\footnote{In the context of eternal black holes we have left and right CFTs in thermofield bouble state. correlation functions like \eqref{OTOCN} are for operators all of which are either on left or on the right CFT. Other probes of chaos are $  \langle W_R(t)   V_L(0) V_R(0) W_R(t)\rangle_{\beta, \mu}$ and $  \langle W_R(t)   V_L(0) V_R(0) W_L(t)\rangle_{\beta, \mu}$. All these correction functions has same ordering for Euclidean time and they are related to each other by analytical continuation.  }
\begin{equation}\label{OTOCN}
	\begin{aligned}
		\langle W(t)V(0)W(t)V(0) \rangle_{\beta, \mu}
	\end{aligned}
\end{equation}
where subscript indicates the particular thermodynamic ensemble under study. $W,V$ are assumed to be approximately local operators (which change energy of the thermal system by an amount of order $1/\beta$) with vanishing one point function. A basic definition of quantum chaos is that correlation functions like \eqref{OTOCN} becomes small at large times (compared to thermal scale) for any choice of $V,W$ \footnote{In certain non-chaotic systems this behaviour might be true for very specific choice of $V,W$ but not for all of them. For example in Ising CFT with primary operators identity, 'spin' and 'energy' only OTOC involving four 'spin' operators vanish at large times. }. Roughly the intuition is that \eqref{OTOC} can be considered as inner product of two states: one in which $V(0)$ acts on thermal state and then $W(t)$, and in the other state  $W(t)$ acts on thermal state first followed by $V(0)$. For large times in a chaotic system these states are rather different and so their inner product is small. This is related to the fact that in a chaotic system at large times commutator squared of the operators \eqref{CC} becomes large, i.e., order $ 2 \langle W(0) W(0) \rangle_{\beta, \mu} \langle V(0) V(0) \rangle_{\beta, \mu}$.\footnote{Correction functions like $  \langle W(t) W(t)  V(0) V(0) \rangle_{\beta, \mu}$ or $  \langle V(0) W(t) W(t)  V(0)  \rangle_{\beta, \mu}$ in large times approach $  \langle W(0) W(0) \rangle_{\beta, \mu} \langle V(0) V(0) \rangle_{\beta, \mu}$. When we expand the commutator squared in \eqref{CC} we get two time ordered and two out of time ordered correction functions. Both the OTOC are related to \eqref{OTOCN}.}

In generic quantum field theory out of time ordered correlation functions  like \eqref{OTOC} requires UV regularization. A convenient way of regularizing it is to define it as (for discussion related to regularization dependance of chaos exponent see \cite{Romero-Bermudez:2019vej})
\begin{equation}\label{DefinitionF}
	\begin{aligned}
		F(t)=\tr \left[y z W(t)y z V(0)y z W(t)y z V(0) \right]
	\end{aligned}
\end{equation}
where we have broken the density matrix into four pieces according to the following definition
\begin{equation}\label{DefinitionYZ}
	\begin{aligned}
	   y^4=\frac{e^{-\beta H}}{Z}, ~~~ z^4=e^{\beta \mu Q}
	\end{aligned}
\end{equation}
With this definition at hand we conjecture the following.

\begin{con}

Consider the out of time ordered correlation function defined in \eqref{DefinitionF} for operators which could create states of arbitrarily large values of charge $Q$. If the function $F$ is well behaved for $t \geq 0$ and the Hilbert space of the system satisfies \eqref{condition} for some fixed constant $\mu_c$, then 
	\begin{equation}\label{FinalEquation}
	\begin{aligned}
	 \frac{1}{1-f(t')} \left| \frac{df(t')}{dt'} \right| \leq \frac{2\pi}{\beta_{eff}}+\mathcal{O}(e^{-\frac{4 \pi}{\beta_{eff}}t'}), ~~ \beta_{eff}=\beta \left(1- \left| \frac{\mu}{\mu_c} \right| \right),
	\end{aligned}
\end{equation}
where $f(t')=F(t_d+t')/F_{*}$ is a ratio of out of time ordered correlation function by a suitable time independent factor. We expect this formula to be true for times (all other coordinates of insertion is assumed to be kept fixed in this process) near scrambling time $0 \ll t' < t_*-t_d $ and  for small values of $\mu/\mu_c \ll 1$ (at temperatures much higher than the scale of any compact dimension if present). 
\end{con}

\subsection{Argument}\label{argument}
In this subsection we present an argument for the conjecture following footsteps of \cite{Maldacena:2015waa}.
Our first  strategy would be to extend the definition of the function $F$ on complex time plane.\footnote{For simplicity the discussion here is restricted to $0+1$ dimensional field theories. In higher space-time dimensions we have extra space co-ordinates (which could be compact or non-compact) to deal with. Nevertheless discussion here can still be applied provided we assume all other coordinates of the operator insertions are kept fixed, in particular they don't depend on time in any particular way (On the other hand it is very important to note that if we scale any other coordinate with time as we focus on large time behaviour of $F$ as defined in \eqref{DefinitionF}, analytic properties of $F$ on complex time plane might change leading to a new bound. An example of this class would be to consider insertion of moving (boosted) operators in a certain non-compact direction. In this situation we must consider the non-compact coordinate as function of time to discuss analytic properties. If all the operators are moving with the same speed we could as well go to their rest frame and apply analysis presented here). When all other directions are non-compact discussion in this section can reliably be applied to late times. If in addition we have a compact dimension of radius $R$, discussion here applies to late times only at high temperatures $\beta \ll R$.}
To achieve this goal we start by defining
\begin{equation}\label{DefFcom}
	\begin{aligned}
				F(t+i \tau)=\tr \left[y z W(t+i \tau)y z V(0)y z W(t+i \tau)y z V(0) \right] ~~~~~~~~~ (t, \tau \in \mathbb{R} )
	\end{aligned}
\end{equation}
We will use following convention for time evolution of operators in Lorentzian time $t$
\begin{equation}
	\begin{aligned}
		W(t)=e^{i H t}W(0)e^{-i Ht}
	\end{aligned}
\end{equation}
Using this we can rewrite \eqref{DefFcom} as
\begin{equation}\label{DefFcom1}
	\begin{aligned}
				F(t+i \tau)= \frac{1}{Z} \tr \left[ e^{\frac{\beta}{4} (\mu Q-H)-\tau H} W(t) e^{\frac{\beta}{4} (\mu Q-H)+\tau H}  V(0) e^{\frac{\beta}{4} (\mu Q-H)-\tau H}  W(t) e^{\frac{\beta}{4} (\mu Q-H)+\tau H}  V(0) \right]
	\end{aligned}
\end{equation}
To discuss analyticity properties of \eqref{DefFcom1} we insert this complete set of states between each two operators, to obtain
\begin{equation}\label{DefFcom2}
	\begin{aligned}
				F(t+i \tau)=& \frac{1}{Z} \sum_{n,m,k,l} e^{-\frac{\beta}{4} (E_n-\mu Q_n)-\tau E_n} W_{n,m}(t) e^{-\frac{\beta}{4} (E_m-\mu Q_m)+\tau E_m}  V_{m,k}(0)  \\
				& ~~~~~~~~~~~ e^{-\frac{\beta}{4} (E_k-\mu Q_k)-\tau E_k}  W_{k,l}(t) e^{-\frac{\beta}{4} (E_l-\mu Q_l)+\tau E_l}  V_{l,n}(0) 
	\end{aligned}
\end{equation}
Here we have used following notation to denote matrix elements of operators between two states
\begin{equation}
	U_{m,n}=\langle \phi_m | U | \phi_n \rangle
\end{equation}

From above form it seems useful to define a new operator $\tilde{H}=H-\mu Q$. For $\mu < \mu_c$, eigen values of $\tilde{H}$ are positive except for finite number of exceptions (since a finite number of exceptions cannot create divergence in the sum \eqref{partitionFun1}). We rewrite \eqref{DefFcom2} in the following form
\begin{equation}\label{DefFcom4}
	\begin{aligned}
				F(t+i \tau)=& \frac{1}{Z} \sum_{n,m,k,l} e^{-\frac{\beta}{4} \tilde{E}_n-\tau \tilde{E}_n-\tau \mu Q_n} W_{n,m}(t) e^{-\frac{\beta}{4}  \tilde{E}_m+\tau \tilde{E}_m+\tau \mu Q_m}  V_{m,k}(0)  \\
				& ~~~~~~~~~~~ e^{-\frac{\beta}{4}  \tilde{E}_k-\tau \tilde{E}_k-\tau \mu Q_k}  W_{k,l}(t) e^{-\frac{\beta}{4}  \tilde{E}_l+\tau \tilde{E}_l+\tau \mu Q_l}  V_{l,n}(0) 
	\end{aligned}
\end{equation}
If any of operators $W,V$ is such that it could create/destroy excitation up to  some  fixed charge, i.e., $ W_{m,n} \neq 0$ or $ V_{m,n} \neq 0$ only for $|Q_m-Q_n|\leq C$ for some fixed constant $C$, in \eqref{DefFcom4} we could simplify factors involving $Q$ to give a term which does not grow with intermediate energy. In this case analytic domain as a function of $\tau$ remains the same as that for $\mu=0$ \footnote{We thank J. Maldacena for a valuable discussion emphasizing this.}. As a result rest of the discussion of this section becomes identical to \cite{Maldacena:2015waa} and maximal chaos exponent in this situation is independent of chemical potential and given by
\begin{equation}
 \lambda_L \leq \frac{2\pi}{\beta}
\end{equation}
To proceed further we restrict ourselves to the situation where $  W_{m,n} \neq 0, V_{m,n} \neq 0$ for arbitrary large values of $Q_m-Q_n$\footnote{One possible example for such a case would be to consider Q as translation generator and considering correlation functions of operators local in position.}. Further we assume these matrix elements does  not grow exponentially in energy for asymptotically large values of energy, more precisely\footnote{We take a moment to understand this assumption little better. For systems satisfying ETH \cite{1999JPhA...32.1163S} matrix elements of a local operator with vanishing one point functions that we are considering here in energy basis satisfies
\begin{equation}
 \begin{aligned}
  X_{m,n}=e^{-S(E^A_{m,n})/2}\tilde{X}(E^A_{m,n},E^D_{m,n})_{m,n}, \ \ E^A_{m,n}=\frac{E_m+E_n}{2}, \ \  E^D_{m,n}=E_m-E_n
 \end{aligned}
\end{equation}
Here $e^{S(E)}$ is the density of states at energy $E$ and  $\tilde{X}$ is expected not to grow exponentially with with energy. This is expected to hold for eigen states in suitable high energy sector. An example would be to consider a conformal field theory on a cylinder of radius $R$ in $d$ space time dimensions. Energy density on the cylinder of an operator of dimension $\Delta$ is $\epsilon=\frac{\Delta}{R^d}$. This suggests following thermodynamic limit where we expect ETH to hold: $R \to \infty$ while holding $\epsilon$ fixed, i.e., $\Delta \sim\epsilon R^d$.
If there exists a similar version of ETH for systems with a global symmetry,   \eqref{assumption} is satisfied when we consider contributions from every heavy energy level. On the other hand if the equation in \eqref{assumption} is true for for any $\epsilon>-\epsilon_*$ with $\epsilon_*>0$ there will be to stronger bound on chaos exponent compared to what we have derived here.  We thank S. Trivedi and H. T. Lam for a discuss related to this.}
\begin{equation}\label{assumption}
	\begin{aligned}
		\lim_{E_n \to \infty} \sum_{E_n \ level} W_{m,n} V_{n,k} \ e^{-\epsilon E_n } =0 ~ ~ ~ \textit{For any $\epsilon>0$} ~ ~ ~ ~ ~ ~ ~ ~ ~ ~  X=V,W 
	\end{aligned}
\end{equation}
Both $H,Q$ are unbounded operators, in this case following ensures convergence of the sum
\begin{equation}\label{TheIneq}
	\begin{aligned}
		\frac{\beta}{4} (|\mu|| Q_n|-E_n)+|\tau| E_n < 0 ~~~~~~ \forall ~  n \sim \Lambda
	\end{aligned}
\end{equation}
Where $\Lambda$ is some UV cut-off scale. Here we have further made the assumption that spectrum is charge conjugation symmetric for this asymptotically large values of energy.\footnote{Otherwise we could get a stronger result, that we don't discuss here.}At this point we restrict ourselves to those class of systems for which
\begin{equation}
		\begin{aligned}
			E_n \geq \mu_c |Q_n| ~~~~~~ \forall ~  n \sim \Lambda
		\end{aligned}
	\end{equation}
It follows that $F(z)$ is analytic in half strip  defined by\footnote{Symmetry of the domain around real axis is a consequence of Schwarz Reflection Principle. }
\begin{equation}\label{domAna}
	\begin{aligned}
		D=\left(z=t+i \tau: t>0, |\tau| < \frac{\beta_{eff}}{4} \right), ~~~~~~ \beta_{eff}=\beta \left(1- \left| \frac{\mu}{\mu_c} \right|
\right)
	\end{aligned}
\end{equation}
Now we turn to bound the function \eqref{DefFcom4} inside the domain given by \eqref{domAna}. We will do so by using unitarity of the underlying Hilbert space. We will use Cauchy-Schwarz lemma on operators defined on the Hilbert space to obtain the bound. At this point  we remind the reader that on the space of matrices following is a positive definite inner product
\begin{equation}
 \begin{aligned}
  (A(t),B(t)) \equiv \tr[A(t)^{\dagger}B(t)]
 \end{aligned}
\end{equation}
This allows us to use  Cauchy-Schwarz inequality
\begin{equation}\label{ChSw}
	\begin{aligned}
		|(A(t),B(t)) | \leq \sqrt{ (A(t),A(t))(B(t),B(t))  }
	\end{aligned}
\end{equation}
To do so we will first express $F$ as inner product of two operators\footnote{Second line is related to the first one through hermitian conjugate.} 
\begin{equation}
	\begin{aligned}
				F(t+i \tau) & = \frac{1}{Z} \tr \left[ y_-^2 W(t)y_+^2 V(0) y_-^2 W(t)y_+^2 V(0) \right] \equiv (A_+(t,\tau),B_+(t,\tau))\\
				&=\frac{1}{Z} \tr \left[ V(0)y_+^2 W(t) y_-^2 V(0)y_+^2 W(t)  y_-^2\right] \equiv (A_-(t,\tau),B_-(t,\tau))
	\end{aligned}
\end{equation}
where we have defined
\begin{equation}
	\begin{aligned}
		& y_{\pm}^2= \frac{1}{Z^{1/4}}e^{\frac{\beta}{4} (\mu Q-H)\pm \tau H} \\
		& A_+(t,\tau)=(y_-W(t)y_+^2V(0)y_-)^{\dagger}=B_+(t,\tau)^{\dagger}\\
		& A_-(t,\tau)=(y_+W(t)y_-^2V(0)y_+)^{\dagger}=B_-(t,\tau)^{\dagger}
	\end{aligned}
\end{equation}
Now we compute norm of these operators 
\begin{equation}\label{norm}
	\begin{aligned}
	&	(A(t)_+,A(t)_+)=(B(t)_+,B(t)_+)=\tr[ W(t) y_-^2 W(t) y_+^2 V(0) y_-^2 V(0) y_+^2]\\
		& (A(t)_-,A(t)_-)=(B(t)_-,B(t)_-)=\tr[ W(t) y_+^2 W(t) y_-^2 V(0) y_+^2 V(0) y_-^2]
	\end{aligned}
\end{equation}
Using Cauchy-Schwarz inequality \eqref{ChSw} we obtain a bound on $F$ by time ordered correction functions. For times larger compared to decay time time-ordered correlation functions factorize due to cluster decomposition principle. Note that for times $t_r$ much larger than scrambling time due to Poincare recurrences this factorization might fail. It follows that for $t_{r}>>t \geq t_d$
\begin{equation}\label{inequilityM}
	\begin{aligned}
		|F(t+i \tau)| \leq F_d^{\pm}(\tau)+\tilde{\epsilon}
	\end{aligned}
\end{equation}
Where we have defined following disconnected correction function 
\begin{equation}
	\begin{aligned}
		F_d^{\pm}(\tau)=\tr[ W(0) y_\mp^2 W(0) y_\pm^2  y_\mp^2  y_\pm^2] \tr[ V(0) y_\mp^2 V(0) y_\pm^2  y_\mp^2  y_\pm^2]
	\end{aligned}
\end{equation}
Here $\tilde{\epsilon}$ denotes possible source of error to factorization of time ordered correlation function. Clearly in the above mentioned time zone $|F(t+i\tau)|$ is bounded by the maximum value of $|F_d^{\pm}(\tau)|$ as $\tau$ varies over \eqref{domAna}.

 Now we will use the fact that given a bounded analytic function we can bound its derive inside the domain by using Schwarz-Pick lemma (see appendix \ref{appendix} for a review). To this end we note that 
\begin{equation}
	\begin{aligned}
		\tr[ X y_\mp^2 X y_\pm^2  y_\mp^2  y_\pm^2] = \tr[(y_\mp X y_\pm^2  y_\mp)^\dagger (y_\mp X y_\pm^2  y_\mp)] \geq 0
	\end{aligned}
\end{equation}
Assuming generically norm of such a state to be non-zero we consider the function 
\begin{equation}\label{fpmP}
 \begin{aligned}
  f^{\pm}(t+i\tau)=\frac{F(t_d+t+i\tau)}{F_d^{\pm}(\tau)}
 \end{aligned}
\end{equation}
This function is real for $\tau=0$ owing to Hermitian nature of $V,W$. It follows from \eqref{inequilityM}, for $t > t_d$, up to errors of order $\tilde{\epsilon}$
\begin{equation}\label{basicIneq}
 \begin{aligned}
  |f^{\pm}(t+i\tau)|\leq 1
 \end{aligned}
\end{equation}
Because of the factor $F_d^{\pm}$ in \eqref{fpmP},  $f^\pm$ is not analytic in the domain given in \eqref{domAna} and hence we cannot use theorem \ref{MathTh}. In what follows with additional assumptions we will try to bound the out of time ordered correction function $F$ by a function which in some sense is close to its factorized value
\begin{equation}
 \begin{aligned}
  F_f=\tr \left[(y z)^2 W(0)(y z)^2 W(0) \right] \tr \left[(y z)^2 V(0)(y z)^2 V(0) \right]
 \end{aligned}
\end{equation}
To this end we note that on the edges of the strip  we have\footnote{For completeness we also report that on the complimentary edges value of $F_d$ are given by $$ F_d^{+} \left(- \frac{\beta_{eff}}{4} \right) =  F_d^{-} \left( \frac{\beta_{eff}}{4} \right) =\tr[ W(0)(xz)^{-1}  W(0)  (yz)^4   (xz) ] \tr[ V(0)(xz)^{-1}  V(0)  (yz)^4   (xz) ]$$  }
\begin{equation}\label{edge}
 \begin{aligned}
  & F_d^{+} \left( \frac{\beta_{eff}}{4} \right) =  F_d^{-} \left( -\frac{\beta_{eff}}{4} \right) \equiv F_{d,s}=\langle W(0) W(0) \rangle_*\langle V(0) V(0) \rangle_* 
 \end{aligned}
\end{equation}
where we have defined 
\begin{equation}\label{DefinitionX}
	\langle X(0) X(0) \rangle_*=\tr[    (xz) X(0)(xz)^{-1} (yz)^2 X(0)  (yz)^2 ],  ~~~ x^4=e^{-\beta \frac{\mu}{\mu_c}H} 
\end{equation}
It follows from \eqref{edge} and \eqref{basicIneq} that the function $f_{s}(t'+i\tau)=F(t_0+t'+i\tau)/F_{d,s}$
is bounded on the edges as $|f_s(t\pm i \beta_{eff})|\leq 1$ upto errors of order $\tilde{\epsilon}$. 

In situations where  $|F_{d,s}|\leq|F_f|$, function $f_{f}(t'+i\tau)=F(t_0+t'+i\tau)/F_{f}$
is bounded on the edges as $|f_f(t\pm i \beta_{eff})|\leq 1$ upto errors of order $\tilde{\epsilon}$. Now we will assume existence of  a time $t_0$ between dissipation and scrambling time $t_d\leq t_0 \leq t_*$ such that 
\begin{equation}\label{leftV}
 \begin{aligned}
  |F(t_0+i \tau)|\leq F_{f}
 \end{aligned}
\end{equation}
In equation \eqref{leftV} there are two possible sources of errors - error in approximating out of time ordered correlation function by time ordered one $\epsilon_1$, error in approximating time ordered correlation function with factorized value $\tilde{\epsilon}$. Below we will come back to these issues in detail and try to argue that in certain systems for small enough values of chemical potential we can choose $t_0$ near scrambling time for this equation to remain valid. With these assumptions in mind the function $f_f $ satisfies corollary \ref{MathTh}. Therefore for $t_d<t_0 < t'+t_d < t_*$ 
	\begin{equation}\label{FinalEquation}
	\begin{aligned}
	 \frac{1}{1-f_f(t')} \left| \frac{df_f(t')}{dt'} \right| \leq \frac{2\pi}{\beta_{eff}}
	\end{aligned}
\end{equation}
Below we make some additional assumptions about detailed form of $f$, this equation remains valid (as long as errors above remain small) even in situations where assumptions below do not stand.

Now we focus our discussion for times scales in between dissipation time and scrambling time $t_0 < t<t_*$. Discussion in this section has been focused to the situation all other coordinates $\vec{x}$ of insertions are kept fixed as we discuss time  
dependence of of the out of time ordered correction function $F$. In other words we have two class of coordinates $t,\vec{x}/t$ such that former is taken large while keeping the later small. In some systems it is possible to assume following simple ansatz for the unknown function $F$ to the leading order in $\vec{x}/t$ expansion\footnote{Actual functional form of the function might be much more complicated than the form presented in \eqref{ansatz1}, as long as we can approximate the function in an interval of time keeping the errors small, we can apply this argument to get a bound on chaos exponent in that interval of time. For times outside this interval separate discussion is needed. In particular if one is interested in discussion of time averaged (over intervals larger compared to the one discussed) growth, this discussion does not apply.}
\begin{equation}\label{ansatz1}
 \begin{aligned}
  \frac{F_{f}-F(t)}{F_{f}}=\epsilon e^{ \lambda_L t}
 \end{aligned}
\end{equation}
here $\epsilon$ is an infinitesimal positive parameter that depends on the system under consideration (in general we have no concrete justification for this particular form of $F$).
 Plugging this form in  \eqref{FinalEquation} we get
\begin{equation}\label{Bound}
 \lambda_L \leq \frac{2\pi}{ \beta \left(1- \left| \frac{\mu}{\mu_c} \right| \right)}
\end{equation}

Now we turn to a discussion of various errors. As a prototype model we will assume two different operators $V,W$ in a large $N$ field theory. Due to large $N$ factorization possible source of error in factorization of time ordered correlation function is $\tilde{\epsilon}=e^{-t/t_d}+\mathcal{O}(N^{-2})$, where the first factor comes from exponential decay of off diagonal factorized part. Whereas we are trying to understand effect of chaotic dynamics   which is order $\epsilon e^{\lambda_L t}$ and we want this to be dominant over $\tilde{\epsilon}$. Clearly this is possible as long as we choose $t_0 \geq \frac{t_*}{1+\frac{1}{\lambda_L t_d}}$. 

Systems that do not fall under the discussion that we had above satisfy $|F_{d,s}|\geq|F_f|$. We need to take a more restrictive path to obtain a bound on chaos exponent in this situation. Under assumptions similar to \eqref{leftV} with $F_f$ replaced by  $F_{d,s}$ we obtain \eqref{FinalEquation} replacing $f_f$ by $f_{s}$. Additional progress can be made in situations where $\epsilon_2=|(F_{d,s}-F_f)/F_f|< \epsilon e^{ \lambda_L t}$, we can approximate \eqref{ansatz1} by
\begin{equation}\label{ansatz2}
 \begin{aligned}
  \frac{F_{d,s}-F(t)}{F_{d,s}}=\epsilon e^{ \lambda_L t}
 \end{aligned}
\end{equation}
At $\mu=0$ we have $F_f=F_{d,s}$, therefore for small values of $\frac{\mu}{\mu_c} \ll 1$, choosing $t_0$ near scrambling time it  be possible  that we can satisfy the required condition.\footnote{Within the framework of finite temperature Schwinger-Keldysh formalism for correlation functions  of operators lying only on one edge of a time-fold it is often the case that they are independent of the contour of integration. A diagrammatic proof can be found in \cite{Matsumoto:1982ry, Matsumoto:1984au}. 
That being said in general ensembles like the ones we are considering here one need to revisit those proofs to see how small $\epsilon_2$ is in general (since it contains insertions on two edges of a single time fold).}  The bound on chaos exponent as given in \eqref{Bound} remains unchanged.

The formula \eqref{Bound} is derived under the assumption that we have a time window of exponential growth more precisely $t_* > t_d$. Therefore this formula certainly stops making sense for $\mu \sim \mu_c \left(1-\frac{1}{\log(\epsilon^{-1})} \right) $. Further we assumed certain simplifying assumptions like  \eqref{ansatz1} or \eqref{ansatz2}, hence equation \eqref{Bound} is valid only under those possible asumptions. Although we didn't require it for the argument presented here in a generic chaotic system for times larger than scrambling time, out of time ordered correction function $F$ is expected  to decay exponentially. 

Here we emphasize situations where the conjecture is not applicable. We have used unitarity and cluster decomposition quite heavily. If any one of these assumptions are not satisfied the conjecture will fail to be true. 
 Also we have restricted the discussion to systems for which scrambling time is much larger compared to decay time. In two dimensional conformal field theories one could easily find situations where both the two point function and the commutator squared becomes significantly at the same time scale and as a result in those situations discussion above does not apply.

\section{Examples}\label{examples}

\subsection{Internal symmetries}

In this subsection we consider generic conformal field theories on cylinder $S^{d-1} \times \mathbb{R}$ in dimensions greater than two $ (d> 2)$.\footnote{We thank X. Yin for discussions related to this.} Energy and internal global symmetry charge density in these theories are given by
\begin{equation}
	\begin{aligned}
		\epsilon=\frac{\Delta_Q/R}{R^{d-1}}, ~~ q=\frac{Q}{R^{d-1}}
	\end{aligned}
\end{equation} 
Where $\Delta_Q$ is the dimension of lightest operator carrying charge $Q$ ($R$ is the radius of the sphere).  Systems in which it is possible to take  $Q$ very large keeping charge and energy density fixed in large charge sector of the theory 
\begin{equation}	\begin{aligned}
		\Delta_Q \sim Q^{\frac{d}{d-1}}
	\end{aligned}
\end{equation}
Detailed analysis of \cite{Hellerman:2015nra, Monin:2016jmo, Jafferis:2017zna} predicts the same result in a generic conformal field theory with a global internal symmetry. For such theories on cylinder we immediately conclude that ($\mu_c \to \infty$) 
\begin{equation}\label{conclusionTrivial}
	\begin{aligned}
		\lambda_L \leq \frac{2 \pi}{\beta} 
	\end{aligned}
\end{equation}
Analysis above excludes theories with a BPS sector - for such theories above limit does not exist. In a supersymmetric theory such BPS operators carry global charges which scale as energy in large charge sector and therefore will generically lead to a finite value of critical chemical potential. A coherent state (when normalisable) obtained for example by exponentiating a charged operator would carry all possible values of the global symmetry under consideration.\footnote{We thank C. Cordova for discussion related to this.} Chaos exponent in correlation function of such operators in strongly coupled regime  could in principle show very interesting dependance on chemical potential. We will not attempt to study these situations in this note. Instead we will turn to discuss spacetime symmetries.

\subsection{Spacetime symmetries}

In this subsection we consider conformal field theories with a holographic dual. We will be considering correlation function of local operators in these theories. First part of the section is devoted to the analysis of two dimensional conformal field theory in large central charge expansion. The we turn to discuss situations when we turn on chemical potential for rotation.

For computational simplicity we restrict ourselves to two spacetime dimensions. In this subsection we closely follow the analysis of \cite{Roberts:2014ifa}.
In a conformal field theory in two spacetime dimensions one can obtain thermal correlation functions starting from zero temperature correlation functions through a conformal mapping. In this section we will exploit this fact and calculate out of time order correlation function of local operators $V,W$ at finite temperature. Keeping the possibilities open we will consider a situation where left  and right moving sectors see different temperatures $T_+, T_-$ respectively.\footnote{This fact is also appreciated in \cite{Turiaci:2016cvo}.}  Locally conformal transformation from thermal cylinder to the plane is given by
\begin{equation}\label{thMap}
	\begin{aligned}
		z^{\pm}(x,t)=e^{ \frac{2\pi}{\beta_{\pm}}( x\pm t) }, ~~~~~~~ \beta_{\pm}=\frac{1}{T_{\pm}}
	\end{aligned}
\end{equation}
Here $t$ is Lorentzian time and $x$ is coordinate along  spatial direction. We adopt the following convention for the metric
\begin{equation}
 \begin{aligned}
  ds^2=dz^+dz^-
 \end{aligned}
\end{equation}

Note that in Euclidean time $\tau=i t$ coordinate $z^{\pm}$ are periodic with period $\beta_{\pm}$ respectively reflecting the effect of non-zero temperature. 
Thermal expectations values in \textit{Euclidean} domain \footnote{Euclidean domain is the configuration of the Lorenzian correlation function when all the points are spacelike separated from each other.}  are related to corresponding expectation values on the plane through the map
\begin{equation}
	\begin{aligned}
		\langle \mathcal{O}(x,t) ..  \rangle_{\beta_\pm} = \left(\frac{2 \pi z^+}{\beta_+}\right)^{h}  \left(\frac{2 \pi z^-}{\beta_-}\right)^{\bar{h}} \langle \mathcal{O}(z^+,z^-) ..  \rangle
	\end{aligned}
\end{equation}
Here $h,\bar{h}$ are holomorphic and anti-holomorphic weights of the operator $\mathcal{O}$. We are interested in the following correlation function 
\begin{equation}\label{OTOC1}
	\begin{aligned}
		\langle W(0,t)V(x,0)W(0,t)V(x,0) \rangle_{\beta_\pm}
	\end{aligned}
\end{equation}
This \textit{Lorentzian} configuration $|t|>|x|$ can be obtained by starting from initial configuration in  \textit{Euclidean} domain $|t|<|x|$ and using suitable analytical continuation that we describe below.\footnote{See \cite{Hartman:2015lfa} and references therein for more detailed description.} 

We consider the following correlation function
\begin{equation}\label{cor}
 \begin{aligned}
  \langle W(z^\pm_1)W(z^\pm_2)V(z^\pm_3)V(z^\pm_4) \rangle=\frac{1}{(z_{12}^+)^{2h_w}(z_{34}^+)^{2h_v} }\frac{1}{(z_{12}^-)^{2\bar{h}_w}(z_{34}^-)^{2\bar{h}_v} }G(z^+,z^-)
 \end{aligned}
\end{equation}
where conformal cross ratios are given by
\begin{equation}\label{crossRatio}
 \begin{aligned}
  z^{\pm}=\frac{z^\pm_{12} z^\pm_{34}}{z^\pm_{13} z^\pm_{24}}
 \end{aligned}
\end{equation}
Following \eqref{thMap} insertion points are given by
\begin{equation}\label{OTOC2}
	\begin{aligned}
		& z^\pm_1= e^{ \frac{2 \pi}{\beta_\pm}(\pm(t'+i \epsilon_1))}
		\\
		& z^\pm_2=e^{ \frac{2 \pi}{\beta_\pm}(\pm(t'+i \epsilon_2))}
		\\
		& z^\pm_3= e^{ \frac{2 \pi}{\beta_\pm}(x\pm i \epsilon_3)}
		\\
		& z^\pm_4= e^{ \frac{2 \pi}{\beta_\pm}(x\pm i \epsilon_4)}
	\end{aligned}
\end{equation}
To obtain \eqref{OTOC1} we start from the Euclidean region and then change $t'$ to $t$ gradually and suitable analytical continuation is dictated by the ordering of the $\epsilon$ s. All the $\epsilon$ s are taken to be infinitesimal maintaining the ordering  \footnote{In the context of eternal black holes, to compute correlation function of operators living on both left and right CFT a calculation with  $\epsilon \sim \beta$ s is required (for example to see the effect of scrambling in a perturbed thermofield double state created by action of a another local operator, see further \cite{Shenker:2013pqa}). As long as this is done maintaining the ordering \eqref{ordering} our conclusion in this section about cross-ratios crossing branch-cut remains unchanged.  }
\begin{equation}\label{ordering}
 \begin{aligned}
  0<\epsilon_1 < \epsilon_3<\epsilon_2<\epsilon_4
 \end{aligned}
\end{equation}

For $|t|>>|x|$ we find
\begin{equation}
 \begin{aligned}
  z^\pm=-e^{ \frac{2 \pi}{\beta_\pm}(\pm x- t')} \epsilon^{\pm}_{12}\epsilon{^\pm}_{34}, ~~~ \epsilon^{\pm}_{ik}=\frac{2 \pi}{\beta_{\pm}}(\epsilon_i-\epsilon_k)
 \end{aligned}
\end{equation}
therefore at large time limit both the cross-ratios go to zero. But this limit is very much different from Euclidean OPE \footnote{In fact there is a way of thinking about his as an OPE limit on Lorentzian cylinder on different sheets.} limit because of behaviour of cross ratios in intermediate values of $t'$ that we will discuss shortly. 

 Note that in both $z^\pm$ numerator in \eqref{crossRatio} is always small whereas denominator can become small in $z^+$ (respectively in $z^-$) only if $t'=x$ (respectively if $t'=-x$). At this critical value of $t'$ from \eqref{crossRatio} it follows that
\begin{equation}\label{lightcone}
 \begin{aligned}
  z^{\pm}|_{t'=\pm x}=\frac{(\epsilon_1-\epsilon_2)(\epsilon_3-\epsilon_4)}{(\epsilon_1-\epsilon_3)(\epsilon_1-\epsilon_4)}>1
 \end{aligned}
\end{equation}
The inequality follows from \eqref{ordering}. 

Considerations of casualty and unitarity imply the function $G$ has brunch points in complex $z^\pm$ plane at $0,1,\infty$ and corresponding branch-cuts are taken here to lie on the real axis in the intervals  $[-\infty,0],[1,+\infty]$. From \eqref{lightcone} it follows that
\textit{As we move to Lorentzian region starting from Euclidean domain $z^{+\sgn{(xt)}}$ crosses the brunch cut starting at $1$ while $z^{-\sgn{(xt)}}$ does not cross such brunch cut. }

The function $f$ appearing in \eqref{cor} can be expanded in terms of contribution from different Virasoro blocks. In a conformal field theory with Einstein gravity dual we expect to keep contribution from multi-stress tensor and their derivatives and no other higher spin block. These contributions are contained in various different blocks of global conformal group $SL(2,\mathbb{R} ) \times SL(2,\mathbb{R} )$ which sum up to give Virasoro block for identity operator exchange. Keeping these considerations in mind in what follows we keep only contribution from Virasoro identity block.

Exact expression of this block is available on the literature. We consider it only in large $c$ limit with $h_v,h_w ,\bar{h}_v,\bar{h}_{w}$ held fixed.\footnote{Note that here we are not making any additional assumption of any one of the operators being heavy contrary to \cite{Roberts:2014ifa, Fitzpatrick:2014vua}.} In this limit exchange of a global primary made out of multi-stress tensors  is $1/c$ suppressed. Leading $1/c $ effect comes from single stress tensor exchange and it is given by \cite{Perlmutter:2015iya}
\begin{equation}
 \begin{aligned}
  & G(z^+,z^-) \sim F_+(z^+)F_-(z^-)\\
  & F_{\pm}(z^\pm) =1+\frac{2 h_v h_w}{c}(z^\pm)^2 \pFq{2}{1}{2,2}{4}{(z^\pm)}+\mathcal{O}\left(\frac{1}{c^2} \right)
 \end{aligned}
\end{equation}

For large times both the cross ratios are very small. From above discussions if $z^+/z^-$ crosses the brunch cut then $z^-/z^+$ does not. So we need to go around in $F_+/F_-$ and then take small cross-ratio approximation whereas for  $F_-/F_+$ we can take small cross ratio approximation directly. To do so explicitly we note that
\begin{equation}
 \begin{aligned}
   \pFq{2}{1}{2,2}{4}{(z)}=\frac{6}{z^3}\left[(z-2)\log(1-z)-2z \right]
 \end{aligned}
\end{equation}
It follows that for $|t|>>|x|$, in a conformal field theory with Einstein gravity dual  we have
\begin{equation}\label{OTOCFinal}
	\begin{aligned}
		\frac{\langle W(0,t+i\epsilon_1 )V(x,0+i\epsilon_3)W(0,t+i\epsilon_2)V(x,0+i\epsilon_4) \rangle_{\beta_\pm}}{ \langle W(0,t+i\epsilon_1)W(0,t+i\epsilon_2)\rangle_{\beta_\pm}  \langle V(x,0+i\epsilon_3)V(x,0+i\epsilon_4) \rangle_{\beta_\pm} } \sim 1-\frac{48 \pi i h_w h_v}{\epsilon^{\pm}_{12}\epsilon^{\pm}_{34}} e^{ \frac{2 \pi}{\beta_\pm}(t\mp x-t_*) }
	\end{aligned}
\end{equation}
Where we have defined scrambling time to be
\begin{equation}
	\begin{aligned}
		t_*=\frac{\beta_{\pm}}{2\pi}\log c
	\end{aligned}
\end{equation}
We remind the reader that in \eqref{OTOCFinal} sign $\pm$ refers to $\sgn(xt)=\pm 1$. 

In RHS of \eqref{OTOCFinal} coefficient of exponential is divergent and origin of this UV divergence is the fact that when $\epsilon$ s are taken to zero two pair of the operators coincide. When we use suitably smeared operators this coefficient is to be understood as proportional to length-scale of smearing and to product of the energies of the excitations created by the operators $V,W$.

Note that a necessary condition for approximations reaching to \eqref{OTOCFinal} to hold is  $\beta<<t-|x| <<  t_*$. This breakdown at larger times is related to Stringy corrections to the correlation function - in other words related to contribution of blocks other than identity Virasoro block.\footnote{Behaviour of a general $SL(2,\mathbb{R})$ primary block of spin $J$ and dimension $\Delta$ after analytic continuation needed for $t>>x>0$ in leading order in large $c$ limit is (here $\beta_\pm=\beta$ assumed)
\begin{equation}
 \begin{aligned}
  e^{\frac{2\pi}{\beta} ( (J-1)t-(\Delta-1)x   )}
 \end{aligned}
\end{equation}
therefore stress tensor block dominates over other primaries if 
\begin{equation}
 \begin{aligned}
  (J-2)t-(\Delta-2)x<0
 \end{aligned}
\end{equation}
 Clearly to justify validity of \eqref{OTOCFinal} certain sparseness condition on operator spectrum  is needed. In fact more careful analysis of OPE coefficients (roughly this determines how operators 'interact' with each other  ) including resummation over higher spin primaries is to be considered as well. For such an analysis in CFT we refer the reader to \cite{Chang:2018nzm}. For analogous analysis from bulk point of view see \cite{Shenker:2014cwa}. On the other hand for rational conformal field theories we expect no chaos since these are integrable. In fact one can compute OTOC in these theories exactly in large time limit in terms of corresponding modular $S$ matrix and quantum dimension of operators, we refer the reader to \cite{Caputa:2016tgt, Gu:2016hoy} for further details.  }

\subsection*{Chemical potential for rotation}

In this subsection we consider holographic conformal field theory living on the boundary of $AdS_3$ whose bulk dual is a rotating BTZ black hole. Temperature and chemical potential for rotation as seen from the boundary is given by
\begin{equation}
 \begin{aligned}
  T=\frac{r_+^2-r_-^2}{2\pi r_+}, ~~~~~~~~ \mu=\Omega_H=\frac{r_-}{r_+}
 \end{aligned}
\end{equation}
Here $r_+,r_-$ are respectively radii of outer and inner event horizon.

From computation of two point function in this background it can be shown that modes rotating in opposite (left) and same (right) direction as the black hole see different temperatures given by
\begin{equation}
 \begin{aligned}
  T_{\pm}=\frac{r_+ \mp r_-}{2 \pi}=\frac{T}{1\pm \Omega_H}
 \end{aligned}
\end{equation}
Using equation \eqref{OTOCFinal} we see (at least for small angular separation) that maximum chaos exponent is seen by modes which are moving in the direction of the black hole and it is given by ($\Omega_H>0$ assumed)
\begin{equation}\label{rotCE}
 \begin{aligned}
\lambda_L=\frac{2\pi}{\beta(1-\Omega_H)}
 \end{aligned}
\end{equation}
To understand this result from the prospective of theorem \ref{TheBound}, we need to consider maximum chemical potential for which the ensemble makes sense.  Following analysis of \cite{Birmingham:2001pj} and considering properties of blackhole micro states contributing to this ensemble we conclude $\mu_c=1$. Therefore comparing with \eqref{TheBound} we note that the bound is saturated. We emphasize that this result is derived  from local analysis in conformal field theory duel to BTZ black hole. Further for the analysis we approximated the cylinder with a plane which works only locally, therefore the analysis can be reliably applied only when $\Omega_H$ is smaller than the radius of the cylinder - in other words equation \eqref{rotCE} is valid for chemical potentials far away from its critical value only. It is valid only at high enough temperature, where BTZ black hole dominates over thermal gas phase, at least for instantaneous growth as a function of time at large enough times (smaller compared to scrambling time).\footnote{Time averaged, over a scale of AdS, behaviour of the out time ordered correction function is not determined from this analysis.} 

From bulk gravitational considerations in BTZ background the same result is first obtained in \cite{Poojary:2018esz} (see also \cite{Jahnke:2019gxr, Mezei:2019dfv}). Here for clarity and completeness we will present a concise discussion. Time dependance of out of time ordered correlation function is determined by considering motion of one probe operator in the background of shockwave created by the other operator and given by
\begin{equation}
	\begin{aligned}
		f(t')=1-a\frac{1}{N^2} e^{\frac{2\pi}{\beta}t'}h(t')+\mathcal{O}(\frac{1}{N^4})
	\end{aligned}
\end{equation} 
Here $a$ is a time independent factor and $h$ is the profile of the localised (transverse angular co-ordinate is localised to a fixed value) shockwave, more explicitly given by
\begin{equation}
	\begin{aligned}
		h(t')=\frac{e^{-\lambda_L^+ (\Omega_Ht' \text{ mod $2 \pi$})}}{1-e^{ -\lambda_L^+ 2\pi} }-\frac{e^{+\lambda_L^- (\Omega_Ht' \text{ mod $2 \pi$})}}{1-e^{+ \lambda_L^- 2\pi} }, ~~ \lambda_L^{\pm}=\frac{2\pi}{\beta(1\pm\Omega_H)}
	\end{aligned}
\end{equation}
To see the fate of the bound we focus on the following quantity
\begin{equation}
	\lambda_L=\frac{1}{1-f(t')} \left| \frac{df(t')}{dt'} \right|=\frac{2\pi}{\beta}+\frac{h'(t')}{h(t')}
\end{equation}
$h$ is a periodic function of time with period $\frac{2\pi}{\Omega_H}$, therefore to understand its effects it is sufficient to understand the behaviour for a whole period. We will be interested in high temperature $\beta<<1$ and small angular velocity of the horizon $\Omega_H<<1 $ with $\Omega_H/\beta=\eta$. To understand the behaviour of $\lambda_L$ we work in the rescaled time co-ordinates $t'=\frac{2\pi}{\Omega_H}\tilde{t}$, to the leading order in $\eta$ 
\begin{equation}
	\begin{aligned}
		\lambda_L=\frac{2\pi}{\beta}-2 \pi \eta \tanh(\frac{2\pi}{\beta}(\pi-2\pi \tilde{t}))+\mathcal{O}(\eta^2)
	\end{aligned}
\end{equation}
Clearly in large temperature limit for first half of the period $\lambda_L =\lambda_L^+$ and for the second half $\lambda_L =\lambda_L^-$. Very importantly we note that the time period for which we can saturate the maximal chaos exponent is dependent on smallness of chemical potential $\Omega_H$.

\section{Concluding remarks}\label{remarks}

In this note we have considered effects of chemical potential for global symmetries on scrambling. In particular we have argued that in certain situations maximum chaos exponent can be dependent on chemical potential. In particular we have seen that effect of presence of non-zero chemical potential may weaken the bound on chaos. In the context of two dimensional conformal field theories we have seen examples of systems that saturate this new chemical potential dependent bound. Analysis in this paper is based on understanding consequences of analyticity of correlation functions. This is by itself a broad subject and demands more investigation - for example on more general spacetime domains (see \cite{Banerjee:2019kjh}) taking into account underlying symmetries of the manifold. 

Although these examples are quintessential, finding examples in higher dimensions (i.e., CFT dimensional higher than two) would be really fascinating.  Study of chaos exponent in presence of a large rotating black holes in such situations is not known in the literature. Nevertheless discussion of such questions at least in the special case of a near extremal black hole (we are considering situations where is region is not super-radiant unstable due to presence of SUSY or otherwise) should be very much possible. Near the horizon region  of such a black hole has a long $AdS_2$ throat described by an effective Jackiw Teitelboim gravity (with possible additional gauge fields). Low energy dynamics near the horizon is captured by the Schwarzian sector living at the end of  $AdS_2$ throat.  \cite{Almheiri:2014cka, Maldacena:2016upp, Nayak:2018qej, Moitra:2018jqs, Moitra:2019bub}.\footnote{We thank S. Minwalla for emphasizing this.} Understanding coupling of probe matter fields to the Schwarzian mode should dictate relevant chaotic dynamics. On the other hand understanding chaotic dynamics on the verge of super-radiance would be extremely interesting in its own right. Another interesting possibility is to turn on chemical potential for translation symmetry and to try to make contact with the works of \cite{Grozdanov:2017ajz, Blake:2017ris, Haehl:2018izb, Blake:2018leo, Grozdanov:2018kkt}.\footnote{Here it is absolutely crucial that we keep in mind conclusions in this paper apply only when we are looking at the correlation function at space distances that are negligible compared to time separation, not when they are comparable.} Can we figure out origin of the modes that contribute to maximal chaos from conformal field theory prospective?

Fast scrambling of black holes has important information theoretic implications.\footnote{The phenomena of chaos has important implications for quantum gravity S-matrix (if exists) which describe processes that are dominated by formation and evaporation of an intermediate black hole. Validity of an effective field theory calculation in presence of the black hole leads to  a relation between a given black hole S-matrix element and another with an additional ingoing particle \cite{Polchinski:2015cea}. This leads to physics of firewall as a consequence of chaotic dynamics.} In the context of eternal black holes, in thermofield double state a region on left CFT with corresponding region in right CFT remains highly locally correlated in $t=0$ boundary time slice let's say. Addition of a quantum at early enough time $|t|$ of the order of scrambling time $t_*$ could break this entanglement completely. Hence as we have argued, for chemical potential near its critical value (where analysis in this paper almost breaks down) scrambling time could (at least naively) decrease significantly (we believed that it might be posible that in a suitable time averaged way effect of this extra exponential growth is washed away) this breakdown of information could take place much earlier compared to the situation of zero chemical potential. It would be interesting to understand this phenomena better - say in the context of \cite{Hayden:2007cs, Hosur:2015ylk, Yoshida:2018ybz}.

\acknowledgments

 Authors of this paper thank H.T. Lam for numerous discussions at various stages of the project. We are grateful for invaluable discussion with J. Maldacena during a visit at IAS Princeton (we thank N. Seiberg for the invitation). We are extremely  thankful to S. Minwalla for intellectual support at every stage of the work. We also thank C. Cordova, A. Gadde, G. Mandal,  S. Raju, R. Loganayagam, S. Shenker, D. Stanford, S. Trivedi and X. Yin for helpful discussions. We thank the DAE, Government of India, for support. We also acknowledge support from the Infosys Endowment for Research on the Quantum Structure of Spacetime. Most of all, we thank the people
of India for generously supporting research.

\appendix

\section{A detour on complex analysis}\label{appendix}

In this section we review few classic results of complex analysis. We will not be presenting proofs of the theorems, they can be found for example in \cite{rudin1976principles, rudin1987real}. 

\textbf{Notations:}
$D(P,r)$, $\bar{D}(P,r)$ are respectively open and closed disc around point $P$ of radius $r$. $U$ is a connected open subset of complex plane.

\begin{thm}\label{MaximumModulus}
	(Maximum modulus) Suppose that $f$ is a holomorphic function on bounded domain $U$, and continuous on $\bar{U}$,\footnote{$\bar{U}$ is the closure of $U$. } then maximum of $|f|$ occur on $\partial U$.\footnote{If $f$ never vanish on $\bar{U}$ then minimum of $|f|$ occur on $\partial U$. }
\end{thm}

\begin{thm}\label{SchwarzPick}
 (Schwarz-Pick) Suppose that $f$ is a holomorphic function on $D(0,1)$, and continuous on $\bar{D}(0,1)$. Say $|f(z)|\leq 1$ on $\partial D(0,1)$, then
 \begin{equation}
  \begin{aligned}
 \abs*{  \frac{f(z_1)-f(z_2)}{1-f(z_1)^*f(z_2)} } \leq \abs*{  \frac{z_1-z_2}{1-z_1^*z_2} }, ~~~~ z_1,z_2 \in D(0,1)
  \end{aligned}
 \end{equation}
\end{thm}
When applied to the special case of $z_1 \to z_2$ gives a bound on the derivate of $f$, more explicitly,
\begin{equation}
 \begin{aligned}
  |f'(z)|\leq \frac{1-|f(z)|^2}{1-|z|^2}
 \end{aligned}
\end{equation}

\subsection*{Analysis on the strip}

Half strip of width $\frac{\beta}{2}$ is defined by $U=\left\{z: Re(z)>0, |Im(z)| < \frac{\beta}{4} \right\}$. Since $U$ is not bounded Maximum modulus principle \ref{MaximumModulus} is not applicable here. Instead a simple generalisation is as follows

\begin{thm}\label{ Phragmen-Lindelof}
	(Phragmen-Lindelof) Consider a function $f: \bar{U} \to \mathbb{C} $ such that $f(z)$ is analytic on $U$ and continuous on $\bar{U}$. Further $|f(z)| \leq 1$ for $z \in \partial U $ and there exists a constant C such that  $|f(z)| \leq C$ on $\bar{U}$ 
then  $|f(z)| \leq 1$ for $z \in  \bar{U} $

\end{thm}

\begin{cor}\label{MathTh}
	Consider a function $f: \bar{U} \to \mathbb{C} $ such that $f(z)$ is analytic on $U$ and continuous on $\bar{U}$. Assume $|f(z)| \leq 1$ for $z \in \partial U $ and there exists a constant C such that  $|f(z)| \leq C$ on $\bar{U}$. Further say $f$ is real valued on the real axis. 
Then growth of the function on positive real axis $z=t$ satisfies the following
\begin{equation}
	\begin{aligned}
	\frac{1}{1-f(t)} \left| \frac{df(t)}{dt} \right| \leq \frac{2\pi}{\beta}+\mathcal{O}(e^{-\frac{4 \pi}{\beta}t})	~~~~~~~~~~~~~~~~(t>0)
	\end{aligned}
\end{equation}
\end{cor}

This can be proved by applying Schwarz-Pick lemma \ref{SchwarzPick} locally after conformally mapping the half strip to the unit disc 
\begin{equation}
 \begin{aligned}
  w=\frac{1-\sinh( \frac{2 \pi}{\beta} z)}{1+\sinh(\frac{2\pi}{\beta}z)}, ~~~~~ z\in U
 \end{aligned}
\end{equation}
For a discussion of this point see \cite{Maldacena:2015waa}.

\textbf{Remark:}
If further the function can be approximated in terms of an infinitesimal parameter $\epsilon>0$ like $f(t)=1-\epsilon e^{\lambda_Lt}+\mathcal{O}(\epsilon^2 )$, then to the leading order in $\epsilon$ above theorem implies $|\lambda_L| \leq \frac{2 \pi}{\beta}$.

\bibliography{cp}

\providecommand{\href}[2]{#2}\begingroup\raggedright

\end{document}